\documentclass[3p,times,twocolumn]{elsarticle}
 \biboptions{comma,sort&compress}
 
\usepackage{graphicx}
\usepackage{amsmath}
\usepackage{dsfont}
\usepackage{here}
\usepackage{ecrc}

\volume{00}

\firstpage{1}

\journalname{Nuclear and Particle Physics Proceedings}

\runauth{}

\jid{nppp}

\jnltitlelogo{Nuclear and Particle Physics Proceedings}

\usepackage{amssymb}

\usepackage[figuresright]{rotating}

\usepackage{color}
\usepackage{ulem}

\def\m{M_{\rm QCD}}

\begin{document}

\begin{frontmatter}

\title{Universal Scale Factors: A Bridge Between Chiral Lagrangians and QCD Sum-Rules 
 $^*$}
 \cortext[cor0]{Talk given at 23rd International Conference in Quantum Chromodynamics (QCD 20),  27--30 October 2020, Montpellier - FR}
 \author[label1]{Amir H.~Fariborz}
\ead{fariboa@sunyit.edu}
\address[label1]{Department of Mathematics/Physics, SUNY Polytechnic Institute, Utica, NY 13502, U.S.A.}
 \author[label2]{J.~Ho
 }
\ead{jason.ho@dordt.edu}
\address[label2]{
Department of Physics, Dordt University, Sioux Center, Iowa, 51250, USA
} 

 \author[label3]{T.G.~Steele\fnref{fn1}}
   \fntext[fn1]{Speaker, Corresponding author.}
    \ead{tom.steele@usask.ca}
\address[label3]{
Department of Physics and
Engineering Physics, University of Saskatchewan, Saskatoon, SK,
S7N 5E2, Canada
}

\pagestyle{myheadings}
\markright{ }
\begin{abstract}
Chiral Lagrangian  mesonic fields can be connected to QCD quark operators via matrix operators containing scale factors.    These scale factor matrices are shown to be constrained by chiral symmetry, resulting in a universal scale factor for each Chiral Lagrangian nonet. QCD sum-rules, combined with mixing angles from Chiral Lagrangian analyses, are used to determine the scale factors for the $a_0$ isotriplet and $K_0^*$ isodoublet scalar mesons.  The resulting scale factors verify the universality property, providing a validation of the scale factor matrices connecting Chiral Lagrangian mesonic fields and quark operators.
\end{abstract}

\end{frontmatter}
The description of light scalar mesons in terms of quark and gluonic constituents provides some of the greatest challenges in hadronic physics   \cite{PDG,Weinberg_13,07_KZ}.  It is generally expected that  mixtures of two-quark, four-quark, and gluonic substructures is necessary to describe the multitude of scalar mesons below 2 GeV \cite{Mec,close,mixing,NR04,06_F,08_tHooft,global,07_FJS4,05_FJS}. 
The strong mixing of quark and  gluonic components is a crucial feature that emerges in many approaches, including Chiral Lagrangians   \cite{close,06_F,oller},
QCD sum-rules \cite{qcdsr_glue_mix1,qcdsr_glue_mix2,qcdsr_glue_mix3,qcdsr_glue_mix4,qcdsr_glue_mix5,GSR_qq_results_mix,narison_review}, 
 and other methods \cite{other_mix1,other_mix2}.  Another important aspect is the inverted hierarchy between two-quark and four-quark states (e.g., Chiral Lagrangians \cite{BFSS2,06_F,global}, bag model \cite{Jaf}  and QCD sum-rules  \cite{Zhang:2000db,Brito:2004tv,Chen:2007xr}).

The focus of this proceedings is   reviewing our work from Refs.~\cite{Fariborz:2015vsa,Fariborz:2019zht,Fariborz:2019vmt}  on the interconnections between Chiral Lagrangian and QCD sum-rule methodologies as applied to the scalar mesons.  Each method is based on a set of key principles: Chiral Lagrangian linear models \cite{NR04,global,07_FJS4,05_FJS}  or non-linear models \cite{Mec,06_F,SS,BFSS2,Blk_rad,RCPT} 
 are founded on chiral symmetry and its breakdown, while QCD sum-rules are founded on QCD-hadron duality  \cite{SVZ,Reinders:1984sr}. Each method faces challenges in application to the scalar mesons: Chiral Lagrangians are oblivious to the underlying four-quark structures (e.g., molecules, diquarks) while QCD sum-rules require parameterization of the broad (and possibly overlapping) scalar states.   In pursuing the interconnections between Chiral Lagrangian and QCD sum-rule methods, we are motivated by the pursuit of synergies leading to new insights and approaches within each method.  Furthermore, we are motivated by the philosophy that features of the scalar sector that  emerge from our analysis bridging the two methodologies are inherently more robust than features emerging from isolated analyses. 
 
In Ref.~\cite{Fariborz:2015vsa} we developed the general framework of scale factor matrices providing the linkage between Chiral Lagrangian mesonic fields and quark-level composite operators (two-quark and four-quark) used in  QCD sum-rules to probe hadronic properties.  Chiral  symmetry was shown to constrain the scale factor matrices to contain one universal scale parameter for each Chiral Lagrangian scalar nonet, and the scale factors were successfully  extracted for the $a_0$ isotriplet sub-system, thereby demonstrating the feasibility of the approach to link Chiral Lagrangian and QCD sum-rules   
\cite{Fariborz:2015vsa}.  Dependence of these extracted scale factors on the non-perturbative gluon condensate QCD sum-rule input parameter as studied in  \cite{Fariborz:2019zht}, allowing an estimate of theoretical uncertainties.  In Ref.~\cite{Fariborz:2019vmt} the crucial universality property was established by demonstrating that  the scale factors extracted from the $K_0^*$ isodoublet   and $a_0$  isoptriplet sub-systems were in excellent numerical agreement.
Other key features establishing the validity of the scale factors linking Chiral Lagrangians and QCD sum-rules included agreement between vacuum expectation values in both approaches as related by the scale factors \cite{Fariborz:2015vsa,Fariborz:2019zht}.

We begin by reviewing the chiral symmetry constraints on the scale-factor matrices connecting Chiral Lagrangian fields and QCD sum-rule operators. 
In the generalized linear sigma model notation of \cite{05_FJS,global} there are two chiral nonets $M$ and $M'$ with identical chiral transformation properties but with different $U_A(1)$ transformations
\begin{gather}
M  \rightarrow U_L \,  M \,  U_R^\dagger\,,  \qquad M\rightarrow e^{2i\nu}M\nonumber \\
M'  \rightarrow U_L \,  M' \,  U_R^\dagger\,, ~\quad  M'\rightarrow e^{-4i\nu}M'~,
\label{M_trans}
\end{gather}
and hence $M$ is associated with two-quark ($\bar q q$) structures and $M'$ with  four-quark ($\bar q q\bar q q$)  structures.   The mesonic nonets contain the bare (un-mixed) chiral Lagrangian mesonic scalar and pseudoscalar  fields
\begin{gather}
M  =  S + i\phi \,,~M'  =  S' + i \phi' 
\label{M_SP}\\
\!\!S=
\begin{pmatrix}
S_1^1 & a_0^+ & \kappa^+  \\
a_0^- & S_2^2 & \kappa^0 \\
\kappa^- & {\bar \kappa}^0 & S_3^3
\end{pmatrix}, 
~
S' =
\begin{pmatrix}
{S'}_1^1 & {a'}_0^+ & {\kappa'}^+  \\
{a'}_0^- & {S'}_2^2 & {\kappa'}^0 \\
{\kappa'}^- & {\bar {\kappa'}}^0 & {S'}_3^3 \\
\end{pmatrix}
\label{SpMES}
\end{gather}
and similarly for the pseudoscalar components in $\phi$ and $\phi'$.  QCD operators that satisfy the chiral transformation properties \eqref{M_trans} for $M$ are 
\begin{equation}
(\m)_a^b = ({\bar q}_R)^b ({q_L})_a  \Rightarrow \left(S_{\rm QCD}\right)_a^b =q_a(x) {\bar q}^b(x)
\end{equation}
where $a$ and $b$ are flavor indices and each can take values of 1 to 3, and there is no loss of generality in the chosen normalization of the QCD operator.
 Similarly,  $M'_{\rm QCD}$ can be  mapped to four-quark composite operators, and the specific forms chosen among the many choices will be outlined below.
 
 The relation between the QCD operators and mesonic fields occurs via scale factor matrices  $I_{M}$ and $I_{M'}$  \cite{Fariborz:2015vsa,Fariborz:2019zht,Fariborz:2019vmt}
 \begin{equation}
M=I_{M} M_{\rm QCD}\,,~M'=I_{M'} M'_{\,\rm QCD}~,
\label{M_scale}
\end{equation}
which have the chiral transformation following properties governed by \eqref{M_trans}:
\begin{gather}
 [U_R,I_{M}]= [U_L,I_{M}]=0~,
  \label{I_M}\\
 [U_R,I_{M'}]= [U_L,I_{M'}]=0~,
 \label{I_Mp}
\end{gather}
and hence the scale factor matrices are multiples of the identity matrix \cite{Fariborz:2015vsa,Fariborz:2019zht,Fariborz:2019vmt}
\begin{equation}
I_M=-\frac{m_q}{\Lambda^3}\times \mathds{1}\,,~I_{M'}=\frac{1}{{\Lambda'}^5}\times  \mathds{1}\, ,
\label{scale_factors}
\end{equation}
where the (constant) scale factor parameters $\Lambda$ and $\Lambda'$ have dimensions of energy and the quark mass factor $m_q=(m_u+m_d)/2$ has been chosen for  renormalization-group purposes as discussed below.  An important consequence emerging from  \eqref{M_scale} and \eqref{scale_factors} is {\em universality}:  the scale factor matrices apply to the entire nonet and hence the scale factors $\Lambda$ and $\Lambda'$ must be identical for all members of the chiral nonets.  Chiral symmetry thus leads to the remarkable result that only two universal scale factors are needed to relate Chiral Lagrangian mesonic fields to QCD operators as expressed in \eqref{M_scale} and \eqref{scale_factors}.   For the remainder of this proceedings, the methodology for determining the scale factors will be presented, and the universality property will be established for the scale factors extracted from the  $K_0^*$ isodoublet   and $a_0$  isoptriplet sectors \cite{Fariborz:2019vmt}.

As an illustration of the methodology, we begin with the $K_0^*$ isodoublet  system and relate the physical $K_0^*(700)$ and $K_0^*(1430)$  chiral Lagrangian fields to QCD operators $J_\kappa^{\rm QCD}$ via a combination of a Chiral Lagrangian rotation matrix $L_\kappa$ and the scale factor matrix $I_\kappa$  \cite{Fariborz:2015vsa,Fariborz:2019zht,Fariborz:2019vmt}: 
\begin{gather}
{\bf K}=
\begin{pmatrix}
K_0^*(700)\\
K_0^*(1430)
\end{pmatrix}
= L_\kappa^{-1}
\begin{pmatrix}
S^3_2\\
\left(S'\right)^3_2
\end{pmatrix}
=
{L_\kappa^{-1} I_\kappa J_\kappa^{\rm QCD} }
\label{K_def}
\\
L^{-1}_\kappa=\begin{pmatrix}
\cos\theta_\kappa & -\sin\theta_\kappa
\\
\sin\theta_\kappa & \cos\theta_\kappa
\end{pmatrix}
\,,~I_\kappa =
\begin{pmatrix}
\frac{-m_q}{\Lambda^3} &0 \\
0 & \frac{1}{{{\Lambda'}^5}}
\end{pmatrix}
~,
\label{I_matrices}
\\
J_\kappa^{\rm QCD}=\begin{pmatrix}
J^{\kappa}_1\\
J^{\kappa}_2
\end{pmatrix}
\,,~
J^{\kappa}_1=\bar ds \\
\begin{split}
J^{\kappa}_2&=\sin(\phi) u^T_\alpha C\gamma_\mu\gamma_5 s_\beta\left(\bar d_\alpha\gamma^\mu\gamma_5 C\bar u_\beta^T-\alpha\leftrightarrow \beta \right)
\\
&
+\cos(\phi) d^T_\alpha C\gamma_\mu s_\beta\left(\bar d_\alpha\gamma^\mu C\bar u_\beta^T+\alpha\leftrightarrow \beta \right)
\end{split}
\end{gather}
where $C$ is the charge conjugation operator and $\cot\phi=1/\sqrt{2}$ \cite{Chen:2007xr}. The analogous expression for the  $a_0(980)$ and $a_0(1450)$ isotriplet system is 
\begin{gather}
{\bf A}=
\begin{pmatrix}
a_0^0(980)\\
a_0^0(1450)
\end{pmatrix}
= L_a^{-1}
\begin{pmatrix}
\frac{S_1^1 - S_2^2}{\sqrt{2}}\\
\frac{{S'}_1^1 - {S'}_2^2}{\sqrt{2}}
\end{pmatrix}
=
L_a^{-1} I_a J_a^{\rm QCD} 
\label{A_def}
\\
L^{-1}_a=\begin{pmatrix}
\cos\theta_a & -\sin\theta_a
\\
\sin\theta_a & \cos\theta_a
\end{pmatrix}
\,,~I_a =I_\kappa=
\begin{pmatrix}
{-m_q\over \Lambda^3} &0 \\
0 & {1\over {{\Lambda'}^5}}
\end{pmatrix}
\\
J_a^{\rm QCD} 
=\begin{pmatrix}
J^a_1\\
J^a_2
\end{pmatrix}\,,~
J^a_1=\left(\bar u u-\bar d d\right)/\sqrt{2}
\label{J_a_QCD}
\\
\begin{split}
J^a_2&=\frac{\sin\phi}{\sqrt{2}}d^T_\alpha C\gamma_\mu\gamma_5 s_\beta\left(\bar d_\alpha\gamma^\mu\gamma_5 C\bar s_\beta^T-\alpha\leftrightarrow \beta \right)
\\
&+\frac{\cos\phi}{\sqrt{2}}d^T_\alpha C\gamma_\mu s_\beta\left(\bar d_\alpha\gamma^\mu C\bar s_\beta^T+\alpha\leftrightarrow \beta \right)
- u\leftrightarrow d\,.
\end{split}
\end{gather}
Because the physical $a_0$ and $K_0^*$ chiral Lagrangian fields occur in Eqs.~\eqref{K_def} and \eqref{A_def} and result in a diagonal hadronic correlation function,
their associated QCD operators define projected physical currents $J^P_s$ that lead to a diagonal QCD physical correlation function matrix $\Pi^P(Q^2)$
\begin{gather}
J_s^P = L_s^{-1} I_s J_s^{\rm QCD}
\,,~s=\{\kappa\,,a\} 
\\
\Pi^P(Q^2) = {\widetilde {\cal T}}^s \Pi^{\rm QCD}(Q^2)  {\cal T}^s\,,~~{\cal T}^s= I_s \, L_s
 \label{phys_corr}
\\
 \Pi^{\rm QCD}_{mn}\left(Q^2=-q^2\right)=\!\!\int\!\!\! d^4x\, e^{iq\cdot x}\langle 0| {\rm T}  \left[ J^{\rm QCD}_m (x) J_n^{\rm QCD}(0)^\dagger \right] |0 \rangle
\end{gather}
where  ${\widetilde {\cal T}}$ denotes the transpose of the matrix ${\cal T}$. Because the physical QCD correlator matrix $\Pi^P(Q^2)$ is diagonal, the off-diagonal correlator term $\Pi^P_{12}(Q^2)$ must satisfy
\begin{equation}
 \Pi_{12}^{\rm QCD} = -
\left[
{
 \frac{    {\widetilde {\cal T}}^\kappa_{11} \Pi_{11}^{\rm QCD} {\cal T}^\kappa_{12}
    + {\widetilde {\cal T}}^\kappa_{12} \Pi_{22}^{\rm QCD} {\cal T}^\kappa_{22}
 }
 {{\widetilde {\cal T}}^\kappa_{11}  {\cal T}^\kappa_{22} + {\widetilde {\cal T}}^\kappa_{12}  {\cal T}^\kappa_{12}
 }
}
\right]~,
\label{constraint}
\end{equation}
which will be used as one of our QCD theoretical inputs.  Similarly, the projected physical correlator can be expressed as a diagonal hadronic correlation function $\Pi^H(Q^2)$ containing the resonance properties and a continuum contribution
\begin{gather}
\Pi^P(Q^2) =\Pi^H(Q^2) \,,~ {\bf H}=\{ {\bf K},~ {\bf A} \}
\\
\begin{split}
\Pi^{\rm H}_{ij}  \left(Q^2 =-q^2\right)
&=\int d^4x\, e^{iq\cdot x}
\langle 0| {\rm T} \left[ {\bf H}_i (x) {\bf H}_j(0) \right] |0 \rangle
\\
&=\delta_{ij} \,
\left(
{1\over {m_{s i}^2 +Q^2-i m_{s i}\Gamma_{si}}}
+ {\rm continuum} \right)\,.
\end{split}
\label{PiH}
\end{gather}
Imposing QCD-hadron duality, the hadronic and QCD contributions to the projected physical correlation functions are equated
\begin{equation}
\Pi^H(Q^2) =\Pi^P(Q^2)= {\widetilde {\cal T}}^s \Pi^{\rm QCD}(Q^2)  {\cal T}^s\,,
\label{correlator_sum_rule}
\end{equation}
and by applying an appropriate transform (e.g., Borel transform \cite{SVZ,Reinders:1984sr}) to both sides of \eqref{correlator_sum_rule} any desired QCD sum-rule can be obtained.  To 
ensure that the rotation matrices $L_s$ properly disentangle the mixture of Chiral Lagrangian mesonic fields and projected physical QCD correlator, Gaussian sum-rules will be used because they can probe multiple states with similar sensitivity  \cite{gauss,harnett_quark}; by contrast Laplace sum-rules will suppress heavier states  and  obscure  incomplete diagonalization.  The resulting relation between the hadronic and projected QCD  Gaussian sum-rules for the $a_0$ ($s=a$) and $K_0^*$ ($s=\kappa$) channels is 
\begin{gather}
G^H\left(\hat s, \tau\right)=\begin{pmatrix}
{G^H}_{11} & 0 \\
0 & {G^H}_{22}
\end{pmatrix}
=\widetilde{\cal T}^s G^{\rm QCD}\left(\hat s,\tau ,s_0\right){\cal T}^s
\label{full_GSR}\,,
\\
G^{H} ({\hat s}, \tau) =\frac{1}{\sqrt{4\pi\tau}}
\int\limits_{s_{th}}^{\infty} \!\! dt \,{\rm exp} \left[  {\frac{-({\hat s} - t)^2}{4\tau}}\right]\,\rho^H(t)\,,
\label{GSR}
\\
\rho^H(t)=
\frac{1}{\pi} {\rm Im} \Pi^{ H}(t) +\theta\left(t-s_0\right){\frac{1}{\pi}} {\rm Im} \Pi^{ \rm QCD}(t)\,,
\label{spectral}
\end{gather} 
where the individual continuum thresholds $s_0^{(1)}$ and $s_0^{(2)}$ for each diagonal entry in the $2\times2$ matrices is collectively denoted by $s_0$. Renormalization group behaviour identifies the Gaussian width $\tau$ as  the renormalization scale and hence $\tau$ has a lower bound from QCD, and we choose $\tau=3\,{\rm GeV^4}$ consistent with the central value used in Refs.~\cite{GSR_qq_results_mix,harnett_quark}.  However, the Gaussian kernel peak  $\hat s$ is unconstrained by QCD and can be varied as a QCD sum-rule parameter analogous to the Borel scale in Laplace sum-rules.

Because the constraint Eq.~\eqref{constraint} is linear in the QCD correlation function, it also applies to the QCD Gaussian sum-rule matrix entries.  Imposing the constraint \eqref{constraint} then provides the Gaussian sum-rules for the diagonal terms in \eqref{full_GSR} for each channel ($s=\{a,\kappa\} $)   \cite{Fariborz:2015vsa,Fariborz:2019zht,Fariborz:2019vmt}
{\allowdisplaybreaks
\begin{gather}
G_{11}^H(\hat s,\tau)=a A
G_{11}^{\rm QCD}\left(\hat s,\tau, s_0^{(1)}\right)-bB
G_{22}^{\rm QCD}\left(\hat s,\tau, s_0^{(1)}\right)
\label{G_eqs}
\\
G_{22}^H(\hat s,\tau)=-aB
G_{11}^{\rm QCD}\left(\hat s,\tau, s_0^{(2)}\right)+bA
G_{22}^{\rm QCD}\left(\hat s,\tau, s_0^{(2)}\right)
\nonumber
\\
A=\frac{\cos^2\theta_s}{\cos^2\theta_s-\sin^2\theta_s}\,,~
B=\frac{\sin^2\theta_s}{\cos^2\theta_s-\sin^2\theta_s}
\\
a=\frac{m_q^2}{\Lambda^6}\,,~b=\frac{1}{\left(\Lambda'\right)^{10}}
\end{gather}
}
where $G_{11}^H$ and $G_{22}^H$ respectively represent $K_0^*(700)$ and  $K^*_0(1430)$ contributions for the $s=\kappa$ isodoublet channel and the $a_0(980)$ and $a_0(1450)$ contributions for the $s=a$ 
isotriplet channel.  The constraint \eqref{constraint} has been examined in \cite{Fariborz:2019zht}, and agrees with the order-of-magnitude estimate for the off-diagonal Gaussian sum-rule $G^{\rm QCD}_{12}$.

Although the constraint \eqref{constraint} has been used in obtaining \eqref{G_eqs}, the solutions for the diagonal elements retain a general property of  \eqref{phys_corr}.  After applying the Borel transform (adapted for Gaussian sum-rules) to  \eqref{phys_corr}, taking the trace,  and noting that $L_s\tilde L_s=\mathds{1}$ so that  ${\cal T}^s {\widetilde {\cal T}}^s=I_s^2$ leads to
\begin{gather}
{\rm Tr}\left[ G^P\right]={\rm Tr}\left[ G^H\right]={\rm Tr}\left[ G^{\rm QCD}I_s^2 \right]
\label{trace_result}
\\
\begin{split}
G_{11}^H(\hat s,\tau)+&G_{22}^H(\hat s,\tau)
\\&
=aG_{11}^{\rm QCD}\left(\hat s,\tau, s_0^{(1)}\right)+bG_{22}^{\rm QCD}\left(\hat s,\tau, s_0^{(1)}\right)\,.
\end{split}
\label{trace_constraint}
\end{gather}
The solutions Eq.~\eqref{G_eqs} that emerge from applying the  diagonalization constraint \eqref{constraint} 
 also satisfy the general property \eqref{trace_constraint} providing a valuable consistency check on our analysis methodology.  However, it is remarkable that Eqs.~\eqref{trace_result} and \eqref{trace_constraint} are independent of the mixing angle matrix $L_s$, and future analyses based on \eqref{trace_result} may yield valuable insights. 

Eq.~\eqref{G_eqs}  can now be solved for the scale factors  $\Lambda$, $\Lambda'$  and the continuum thresholds are optimized to minimize the $\hat s$ dependence of the scale parameters.
For our detailed analysis, the results for the QCD correlation functions are given in Refs.~\cite{Chen:2007xr,GSR_qq_results_mix,harnett_quark,Zhang:2009qb,Du:2004ki} and the methods of \cite{harnett_quark} can then be used to form the Gaussian sum-rules.   For the QCD input parameters we use PDG values \cite{PDG} (quark masses, and $\alpha_s$)
 and the following QCD condensate \cite{Reinders:1984sr,Narison:2011rn,Beneke:1992ba,Belyaev:1982sa} and instanton liquid model parameters \cite{Shuryak:1982qx,Schafer:1996wv}
{\allowdisplaybreaks
\begin{gather}
 \langle\alpha_s G^2\rangle
      = (0.07\pm 0.02)\, {\rm GeV^4} \,,
      \label{GG}
      \\
\frac{\left\langle \overline{q}\sigma G q\right\rangle}{\langle \bar q q\rangle}=\frac{\left\langle \overline{s}\sigma G s\right\rangle}{\langle \bar s s\rangle}=(0.8\pm 0.1) \,{\rm GeV^2}
\label{mix}
\\
\langle \bar q q\rangle=-\left(0.24\pm 0.2 \,{\rm GeV}\right)^3\,,~\langle \bar s s\rangle=(0.8\pm 0.1)\langle \bar q q\rangle
\label{O6}
\\
  n_{{c}} = 8.0\times 10^{-4}\ {\rm GeV^4}~,~\rho =1/600\,{\rm MeV}~,
  \label{inst}
  \\
  m^*_q=170\,{\rm MeV}\,,~m^*_s=220\,{\rm MeV}~.
\end{gather}
}
The instanton parameters $\rho$ and $n_c$ have an estimated uncertainty of 15\% and the quark zero-mode effective masses $m^*$ have a correlated  uncertainty with $\rho$ and the quark condensate \cite{Shuryak:1981ff}. 
The $m_s/m_q=27.3$ ratio \cite{PDG} is particularly significant because it is both a  QCD input and  a chiral Lagrangian parameter.   The mixing angles $\cos\theta_\kappa=0.4161$  and $\cos\theta_a=0.6304$ are provided by Chiral Lagrangian analyses \cite{05_FJS,global} and are self-consistent with our input parameter $m_s/m_q=27.3$.
For the physical mass and width of the $K_0^*$ states we use  the following values \cite{PDG}: $m_\kappa=824 \,{\rm MeV}$ and $\Gamma_\kappa=478\,{\rm MeV}$ for the $K_0^*(700)$, $m_K=1425 \,{\rm MeV}$ and $\Gamma_K=270\,{\rm MeV}$ for the $K_0^*(1430)$, and for the $a_0(980,1450)$  states  $m_{a}=\{0.98, 1.47\}{\,\rm GeV}$ and $\Gamma_a=\{0.06,0.265\}{\,\rm GeV}$.

\begin{figure}[htb]
\centering
\includegraphics[width=\columnwidth]{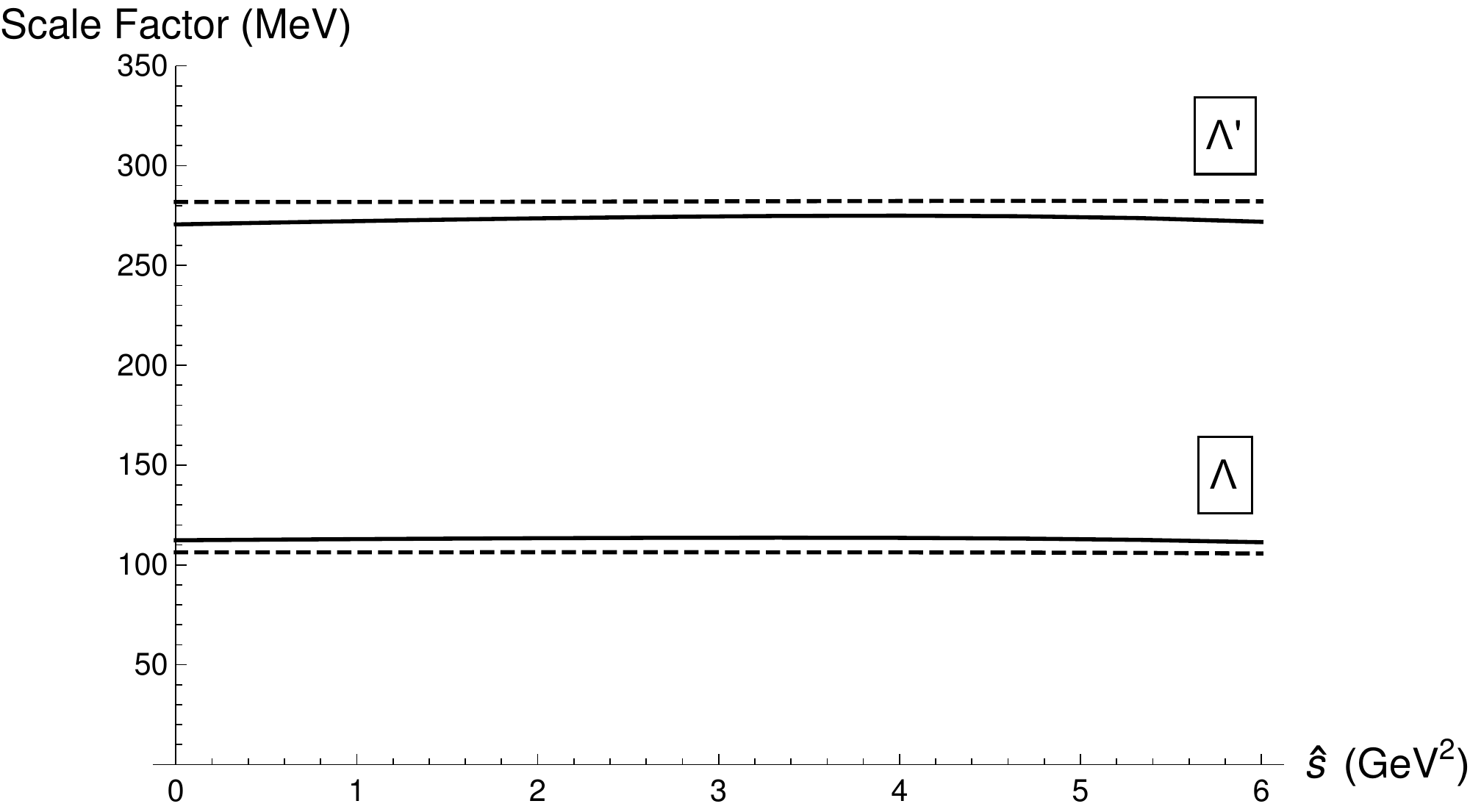}\hspace{0.02\columnwidth}
\caption{The scale factors $\Lambda$ (lower pair of curves) and $\Lambda'$ (upper pair of curves) are shown as a function of $\hat s$
for optimized continuum thresholds in Table \ref{scale_tab}.  Solid curves are for the $K_0^*$ channel and dashed curves are for the $a_0$ channel.
}
\label{scale_fig1}
\end{figure}

The optimized values for the scale factors separately determined for  the $a_0$ and $K_0^*$ channels are shown in Fig.~\ref{scale_fig1} as a function of the Gaussian sum-rule parameter $\hat s$ and their final determinations are given in Table~\ref{scale_tab} \cite{Fariborz:2019vmt}.    Theoretical uncertainties in the scale factors arising from the gluon condensate (the dominant non-perturbative input parameter) have been studied in Ref.~\cite{Fariborz:2019zht}. 
The scale factor results of  Fig.~\ref{scale_fig1} and Table~\ref{scale_tab}  demonstrate the following three important properties that establish the integrity of our proposed connection of Chiral Lagrangian mesonic fields to QCD operators via scalar factor matrices:  
\begin{itemize}
\item Refs.~\cite{Fariborz:2015vsa,Fariborz:2019zht} have examined the relation between the
 Chiral Lagrangian vacuum expectation values  and scale factors  given by
$\langle S_1^1\rangle=-m_q\langle\bar u u\rangle/\Lambda^3$ 
and
  $\langle {S'}_1^1\rangle\approx 1.31\langle \bar d d\rangle \langle \bar s s\rangle/\Lambda'^{\,5}$ (n.b., vacuum saturation effects are embedded in the 1.31 numerical factor), and conclude that any numerical disagreements can be attributed to the known deviations from the vacuum saturation hypothesis.

\item The scale factors have minimal energy ($\hat s$) dependence indicating that no significant additional dynamics is necessary to supplement our fundamental relation \eqref{M_scale} relating Chiral Lagrangian fields and QCD operators. 

\item The scale factors are nearly identical for the $K_0^*$ and $a_0$ subsystems, establishing universality of the scale factors as required by chiral symmetry.  

\end{itemize}
Our future work will continue to seek new insights on the scalar mesons by building upon the universal scale factor matrices that we have firmly established in the isodoublet $K_0^*$ and isotriplet $a_0$ sectors. Of particular interest is the challenging case of the isosinglet $f_0$ system, which in addition to quark-antiquark and four-quark chiral nonets involve the inclusion of scalar and pseudoscalar glueballs. This will further test the universality requirement and allow a more careful examination of the substructure of $f_0$ states.
Informed by the scale factor matrix relations between mesonic fields and QCD operators, 
we will also pursue new methodologies that exploit this synergistic connection.

\begin{table}[htb]
\begin{tabular}{c|cccc}
\hline
\rule{0pt}{3ex}   
Channel &  $s_0^{(1)}$ & $s_0^{(2)}$ & $\Lambda$ & $\Lambda'$ 
\\[2pt]
\hline
$K_0^*$ &1.61 & 3.04 & 0.114 & 0.276 
\\
$a_0$ & 1.68 & 2.88 & 0.106 & 0.282 
\\
\hline
\end{tabular}
\caption{Values for the optimized scale factors $\Lambda,~\Lambda'$ and continuum thresholds 
$s_0^{(1)},~s_0^{(2)}$  for the $a_0$ and $K_0^*$ channels.
All quantities are in appropriate powers of GeV. 
}
\label{scale_tab}
\end{table}

\section*{Acknowledgments}
This research was generously supported in part by the SUNY Polytechnic Institute Research Seed Grant Program.
TGS is grateful for the hospitality of AHF and SUNY Polytechnic Institute while this work was initiated.  
TGS and JH are grateful for research funding from the Natural Sciences and Engineering Research Council of Canada (NSERC), and 
AHF is grateful for a 2019 Seed Grant,  and the support of the College of Arts and Sciences, SUNY Polytechnic Institute.
We thank Research Computing at the University of Saskatchewan for computational resources.

\end{document}